\documentclass[english,aps,manuscript]{revtex4}
\usepackage[T1]{fontenc}
\usepackage[latin9]{inputenc}
\usepackage{bm}
\usepackage{amsmath}
\usepackage{amssymb}
\usepackage{graphicx}
\usepackage{esint}

\makeatletter
\@ifundefined{textcolor}{}
{%
 \definecolor{BLACK}{gray}{0}
 \definecolor{WHITE}{gray}{1}
 \definecolor{RED}{rgb}{1,0,0}
 \definecolor{GREEN}{rgb}{0,1,0}
 \definecolor{BLUE}{rgb}{0,0,1}
 \definecolor{CYAN}{cmyk}{1,0,0,0}
 \definecolor{MAGENTA}{cmyk}{0,1,0,0}
 \definecolor{YELLOW}{cmyk}{0,0,1,0}
}

\usepackage{babel}

\usepackage{babel}

\makeatother

\usepackage{babel}
\begin{document}

\title{Magnetic compressibility and ion-temperature-gradient-driven microinstabilities
in magnetically confined plasmas}

\author{A. Zocco$^{1}$, P. Helander$^{1}$ and J. W. Connor$^{2}$}

\address{$^{1}$Max-Planck-Institut für Plasmaphysik, D-17491, Greifswald,
Germany}

\address{$^{2}$Culham Science Centre, Abingdon, Oxon, OX14 3DB, UK}
\begin{abstract}
The electromagnetic theory of the strongly driven ion-temperature-gradient
(ITG) instability in magnetically confined toroidal plasmas is developed.
Stabilizing and destabilizing effects are identified, and a critical
$\beta_{e}$ (the ratio of the electron to magnetic pressure) for
stabilization of the toroidal branch of the mode is calculated for
magnetic equilibria independent of the coordinate along the magnetic
field. Its scaling is $\beta_{e}\sim L_{Te}/R,$ where $L_{Te}$ is
the characteristic electron temperature gradient length, and $R$
the major radius of the torus. We conjecture that a fast particle
population can cause a similar stabilization due to its contribution
to the equilibrium pressure gradient. For sheared equilibria, the
boundary of marginal stability of the electromagnetic correction to
the electrostatic mode is also given. For a general magnetic equilibrium,
we find a critical length (for electromagnetic stabilization) of the
extent of the unfavourable curvature along the magnetic field. This
is a decreasing function of the local magnetic shear. 
\end{abstract}
\maketitle

\section{Introduction}

Most kinetic investigations of ion-temperature-gradient (ITG) instabilities
in plasmas rely on the simplifying assumption that perturbations are
electrostatic \cite{1.1762151,1.863486,1.859023}. With some notable
exceptions \cite{1.863121,0029-5515-20-11-011,1.860623,1.873449,1.1730294,0741-3335-55-12-125003,0741-3335-56-1-015007},
electromagnetic perturbations have been considered mainly from a numerical
standpoint \cite{1.873694,0029-5515-40-3Y-331,1.3495976,1.3432117,1.4876960},
and the attempt to understand their role in ITG stability has resulted
in a patchy collection of numerical findings rather than in a coherent
physical picture. Moreover, most studies have neglected magnetic compressibility
and have thus neglected the magnetic perturbations parallel to the
equilibrium magnetic field, $\delta B_{\|}$, that are generated by
the instability to maintain perpendicular pressure balance.

From an analytical point of view, the equations describing electromagnetic
ITG modes are present in the works of Antonsen and Lane \cite{1.863121},
and Tang, Connor and Hastie \cite{0029-5515-20-11-011}, where the
linear theory of kinetic ballooning modes was formulated. However,
since the equations derived in these papers are general and thus encompass
many types of instabilities, the role of ITG modes is somewhat obscured.
Later, Kim et al. \cite{1.860623} focused on the physics of the toroidal
ITG instability, extending previous electrostatic work to finite $\beta$
(the ratio of the ion to magnetic pressure) by including the effect
of the induced electric field, $-\partial A_{\|}/\partial t$, on
the electron motion along the magnetic field. Therefore, effects of
$A_{\|}$ (the parallel component of the magnetic vector potential)
were included, however, effects of $\delta B_{\|}$ were neglected
on the grounds that $\beta$ was considered to be small.

A complete electromagnetic theory of ITG modes must retain all three
gyrokinetic fields: $\phi$ (the electrostatic potential), $A_{\|}$
and $\delta B_{\|}$. Formally, the latter two are finite-$\beta$
effects, but as we shall see they become important at surprisingly
low values of $\beta$ because of other small parameters present in
the problem. A general finite-$\beta$ theory must necessarily describe
several families of instabilities, such as ITG Alfvénic modes \cite{1.873449},
$\beta-$induced Alfvénic eigenmodes \cite{983295819921101}, $\beta-$induced
temperature gradient eigenmodes \cite{Miksharapov}, and kinetic \cite{1.863121,0029-5515-20-11-011}
and ideal ballooning modes \cite{CHT}. In this work, we limit ourselves
to the analysis of curvature-driven ITG modes by adopting an ordering
scheme which excludes other instabilities but, at the same time, allows
a small value of $\beta$ ($\beta\ll1$) to affect the ITG mode through
both $A_{\|}$ and $\delta B_{\|}$. The result is a simple formulation
shedding light on why and when electromagnetic effects are important
for toroidal ITG instabilities.

From a numerical point of view, early gyrokinetic simulations \cite{1.873694,0029-5515-40-3Y-331}
had already found magnetic compressibility to be important, in particular
to cancel the stabilizing effect of the ``self-dug'' magnetic well
for drift instabilities \cite{PLA:4728700}. Waltz and Miller reported
on such a cancellation, resulting in a substitution rule for the magnetic
drift: magnetic compressibility could be dropped if the magnetic drift
were replaced by the curvature drift \cite{1.873694}. While this
fact now seems to be common knowledge in part of the gyrokinetic community
\cite{1.3432117,Tobias}, the picture that emerges from systematic
electromagnetic gyrokinetic simulations of the ITG mode is more complicated
\cite{1.3495976} and difficult to disentangle. A simple analytical
explanation is therefore helpful.

In the present work, we build on the recent electrostatic linear theory
of Plunk et al. \cite{1.4868412}, exploit asymptotic techniques to
solve the kinetic problem of the ITG instability, and identify the
conditions that allow this theory to accommodate electromagnetic perturbations.
Somewhat to our surprise, we find that, for strongly driven modes,
magnetic compressibility can be as important as perpendicular magnetic
perturbations for values of $\beta$ accessible to both tokamaks and
stellarators. The ions contribute to magnetic perturbations to maintain
pressure balance, whereas the electrons can have both a stabilizing
and destabilizing effect, depending on the value of $\beta.$ In the
case of a uniform equilibrium magnetic field, a new critical $\beta$
for the electromagnetic stabilization of the toroidal ITG is calculated.
This differs from the one given by Kim et al. \cite{1.860623} in
a fundamental way. A similar stabilization is predicted when an additional
fast particle population is considered. For sheared magnetic equilibria,
the boundary of marginal stability for the electromagnetic component
of the ITG is given, for the first time, using a local approximation
of the magnetic drifts.

\section{Physical Picture}

To understand the role of magnetic perturbations for ITG modes, it
is useful to start with a physical picture of the instability. We
follow Rosenbluth and Longmire, who first described the physical mechanism
responsible for interchange modes \cite{Rosenbluth-Longmire}. The
same description works for the curvature-driven branch of the ITG
modes and will be used here.

Consider a plasma with gradients of the temperature and the magnetic
field strength in the direction of $-\nabla x$. The magnetic field
points in the $z$-direction, and for simplicity we take the density
gradient to vanish. The ion guiding centers drift in the direction
${\bf B}\times\nabla B$, i.e., in the negative $y$-direction, and
do so with a speed that decreases with increasing $x$, since the
drift velocity is proportional to the energy.

\begin{figure}[htb]
\raggedright{}\includegraphics[width=0.7\textwidth]{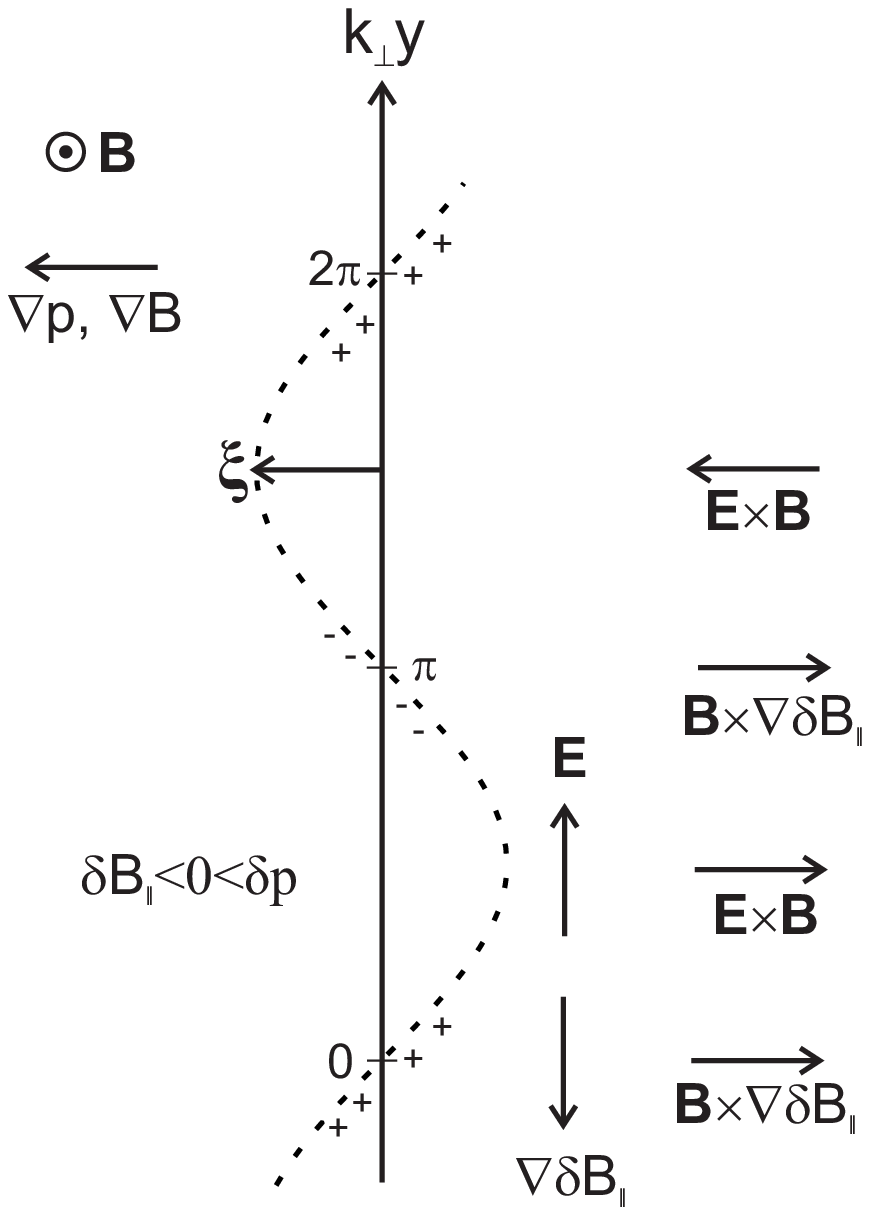}
\caption{Physical mechanism of the curvature-driven ITG instability. In equilibrium,
the ions drift down (in the negative $y$-direction) and the electrons
drift up. A sinusoidal displacement $\mathbf{\xi}$ of the plasma
results in a positive ion pressure perturbation, $\delta p>0$, at
$k_{\perp}y=(2n+1/2)\pi$ and a corresponding negative perturbation
at $k_{\perp}y=(2n-1/2)\pi$. Since the magnetic drift velocity is
proportional to energy, there is an excess of ions drifting downward
where $\delta p_{i}>0$. Ion guiding centers will therefore accumulate
at $k_{\perp}y=2n\pi$ and a corresponding deficit forms at $k_{\perp}y=(2n+1)\pi$,
which creates an upward electric field at $k_{\perp}y=(2n+1/2)\pi$
and an ${\bf E}\times{\bf B}$ drift that reinforces the initial perturbation.
An instatbility thus arises. Furthermore, pressure perturbations are
anticorrelated with perturbations of the magnetic field strength,
$\delta B_{\|}$. The latter therefore cause a horizontal perturbed
${\bf B}\times\nabla B$ drift at $k_{\perp}y=2n\pi$, which enhances
the accumulation of positive charge at these points and strengthens
the instability.}

\label{Fig:fig1} 
\end{figure}

If the plasma is displaced by an ${\bf E}\times{\bf B}$ drift in
the $x$-direction by the distance 
\begin{equation}
\bm{\xi}=\hat{\mathbf{x}}\xi_{0}\sin\left(k_{\perp}y\right),\label{eq:xi}
\end{equation}
the ion pressure is perturbed according to 
\[
\delta p_{i}=-\mathbf{\bm{\xi}}\cdot\nabla p_{i},
\]
where $p_{i}$ is the equilibrium ion plasma pressure. The ion guiding
centers will then start accumulating at $k_{\perp}y=2n\pi$ and a
corresponding deficit of ion guiding centers arises at $k_{\perp}y=(2n+1)\pi$,
see Fig.~\eqref{Fig:fig1}. An electrostatic potential, $\phi=\phi_{0}\cos k_{\perp}y,$
thus builds up (with $\phi_{0}$ having the same sign as $\xi_{0}$)
and gives rise to an ${\bf E}\times{\bf B}$ drift, 
\[
\frac{\partial\bm{\mathbf{\xi}}}{\partial t}=\frac{{\bf b}\times\nabla\phi}{B}=\frac{\phi_{0}}{B}\hat{{\bf z}}\times\nabla\cos k_{\perp}y=\hat{{\bf x}}\frac{k_{\perp}\phi_{0}}{B}\sin k_{\perp}y=\frac{k_{\perp}\phi_{0}}{B}\bm{\mathbf{\xi}},
\]
that amplifies the initial perturbation \eqref{eq:xi}. In this picture
of the instability, the motion of the ions parallel to the magnetic
field is neglected, so it is tacitly assumed that $k_{\|}v_{thi}\ll\omega$,
where $v_{thi}=(2T_{i}/m_{i})^{1/2}$ denotes the ion thermal speed
and $\omega/k_{\|}$ the parallel phase velocity of the instability.
The electrons, on the other hand, can be expected to move quickly
compared with the instability, $k_{\|}v_{the}\gg\omega$, and will
therefore only experience a small ${\bf E}\times{\bf B}$ displacement.

How is this mechanism affected by electromagnetic terms within the
gyrokinetic description of the instability? As already mentioned,
there are two such terms, proportional to $A_{\|}$ and $\delta B_{\|}$
(the perturbation of the magnetic field strength), respectively. The
first one describes the effect of the inductive electric field and
is important to the electrons, which unlike the ions have time to
move significant distances along the magnetic field during the evolution
of the instability. They are therefore sensitive to the parallel electric
field, 
\[
E_{\|}=-\nabla_{\|}\phi-\frac{\partial A_{\|}}{\partial t}.
\]
Instead of $A_{\|}$, we introduce the quantity $\psi$, defined by
\[
\nabla_{\|}\psi=-\frac{\partial A_{\|}}{\partial t},
\]
so that $E_{\|}=-\nabla_{\|}(\phi-\psi)$. Amp\`ere's law, $k_{\perp}^{2}A_{\|}=\mu_{0}J_{\|}$,
then implies 
\begin{equation}
\nabla_{\parallel}^{2}\psi=-\frac{\mu_{0}}{k_{\text{\ensuremath{\perp}\ }}^{2}}\frac{\partial}{\partial t}\nabla_{\parallel}J_{\parallel}\label{eq:Ampere}
\end{equation}
where $\nabla_{\|}J_{\|}$ describes the local accumulation of electrons
due to their parallel motion. If the inductive field is weak, $\psi\ll\phi$,
the electrons are approximately Boltzmann-distributed, 
\[
\frac{\delta n}{n}=\frac{e\phi_{0}}{T_{e}}\cos k_{\perp}y,
\]
so that 
\[
\nabla_{\|}J_{\|}\sim e\frac{\partial\delta n}{\partial t}=\frac{ne^{2}}{T_{e}}\frac{\partial\phi_{0}}{\partial t}\cos k_{\perp}y.
\]
Hence and from Eq.~\eqref{eq:Ampere} we obtain the estimate 
\[
\psi\sim\frac{\mu_{0}ne^{2}}{k_{\perp}^{2}k_{\|}^{2}T_{e}}\frac{\partial^{2}\phi_{0}}{\partial t^{2}}\cos k_{\perp}y,
\]
and we conclude that the critical $\beta=2\mu_{0}nT/B^{2}$ above
which electromagnetic effects are important, $\psi\sim\phi$, scales
as 
\begin{equation}
\beta_{c}\sim\left(\frac{k_{\perp}\rho_{i}k_{\|}c_{s}}{\omega}\right)^{2}.\label{eq:beta1}
\end{equation}
where $c_{s}$ denotes the sound speed, and $\rho_{i}$ the ion Larmor
radius. The ITG mode has a frequency of order $\omega_{\ast}\sim(k_{\perp}\rho_{i})c_{s}/L_{\perp}$,
where $L_{\perp}$ is the length scale of the cross-field gradients,
it thus follows that the critical beta is $\beta_{c}\sim(k_{\|}L_{\perp})^{2}\sim\epsilon^{2},$
and, in a standard tokamak, can be ordered as the square of the inverse
aspect ratio. This is the basic reason why electromagnetic effects
are already important in standard tokamak situations when $\beta\sim10^{-2}$
rather than when $\beta=\mathcal{O}(1)$. This critical beta also
defines the value at which kinetic Alfvén waves are relevant, since
\begin{equation}
\beta_{c}\sim\left(\frac{k_{\perp}\rho_{s}k_{\|}v_{A}}{\omega}\right)^{2}\beta,
\end{equation}
and $\omega=k_{\perp}\rho_{s}k_{\|}v_{A}$ is the kinetic Alfvén wave
dispersion relation.

The other electromagnetic term in the gyrokinetic equation involves
$\delta B_{\|}$ and is sometimes neglected in analytical treatments
and numerical simulations of the gyrokinetic equation. Physically,
it accounts for the perturbation in the $\nabla B$ drift due to the
variation in magnetic field strength, 
\[
\nabla B=\nabla\left|{\bf B}+\delta{\bf B}\right|\simeq\nabla(B+\delta B_{\|}).
\]
The latter is determined by perpendicular pressure balance, 
\[
\delta\left(p_{\perp}+\frac{B^{2}}{2\mu_{0}}\right)=0,
\]
with $p_{\perp}=p_{\perp,i}+p_{\perp,e}$, which implies 
\[
\delta B_{\|}=-\frac{\mu_{0}\delta p_{\perp}}{B}\simeq\frac{\mu_{0}p_{i}'\xi_{0}}{B}\sin k_{\perp}y,
\]
and thus gives rise to a perturbed $\nabla B$-drift of the ions

\begin{equation}
\delta{\bf v}_{di}=\frac{v_{\perp}^{2}}{2\Omega_{i}B}{\bf b}\times\nabla\delta B_{\|}\simeq\underset{I}{\underbrace{-\hat{{\bf x}}\frac{\mu_{0}v_{\perp}^{2}p_{i}'k_{\perp}\xi_{0}}{2\Omega_{i}B^{2}}\cos k_{\perp}y}},\label{eq:vdriftpert}
\end{equation}
where ${\bf b}={\bf B}/B$ and $\Omega_{i}=eB/m_{i}$. As is clear
from Fig. \eqref{Fig:fig1}, this extra drift reinforces the density
accumulation around $k_{\perp}y=2n\pi$ and thus amplifies the instability.
It does so even in the absence of a density gradient, since the $\nabla B$-drift
is proportional to the perpendicular kinetic energy and we assume
that a temperature gradient is present.

There is, however, also a third effect of finite plasma pressure,
since this affects the equilibrium magnetic field by making the curvature
vector deviate from the gradient of the field strength, 
\[
\boldsymbol{\kappa}=\frac{\nabla_{\perp}B}{B}+\frac{\mu_{0}\nabla p}{B^{2}},
\]
where $p=p_{i}+p_{e}.$ The equilibrium ion drift velocity can thus
be written 
\begin{equation}
{\bf v}_{d}=\underset{II}{\underbrace{\left(\frac{v_{\perp}^{2}}{2}+v_{\|}^{2}\right)\frac{{\bf b}\times\boldsymbol{\kappa}}{\Omega_{i}}}}-\frac{\mu_{0}v_{\perp}^{2}}{2\Omega_{i}B^{2}}{\bf b}\times\nabla(\underset{III}{\underbrace{p_{i}}}+\underset{IV}{\underbrace{p_{e}}})={\bf v}_{\kappa}+\Delta{\bf v}_{d},\label{eq:vdrift}
\end{equation}
where the second term on the right ($III+IV$) opposes the basic curvature
drift (term $II$) causing the instability. Thus, if the plasma pressure
is increased whilst the magnetic curvature is kept fixed, then the
drift velocity is reduced and the instability is weakened. As has
been discussed in the literature \cite{Berk,0029-5515-20-11-011,1.3432117,1.3495976},
this effect from the ions partly cancels that from $\delta B_{\|}$,
but it is important to keep in mind that this cancellation only holds
if $\mathbf{\boldsymbol{\kappa}}$, rather than $\nabla_{\perp}B$,
is held constant. A simple mathematical argument for the cancellation
is given in an Appendix.

Even though the electrons contribute relatively little to the ion
instability, their pressure gradient exerts a stabilizing effect.
While term $III$ in Eq. \eqref{eq:vdrift} tends to cancel the perturbed
grad-$B$ drift $I$ of Eq. \eqref{eq:vdriftpert}, the diamagnetic
electron contribution {[}term $IV$ in Eq. \eqref{eq:vdrift}{]},
tends to oppose the drive of the mode $II.$ When these terms balance,
we have 
\begin{equation}
\beta_{e}\sim\frac{L_{p}}{R},\label{eq:betaZHC}
\end{equation}
where we used $\nabla B\sim B/R,$ and $\nabla p_{e}\sim p_{0}/L_{p}.$
This stabilizing influence of finite $\beta$ was studied by Hastie
and Taylor for MHD instabilities in a combined mirror-cusp magnetic
configuration \cite{JimBryan}, and by Rosenbluth and Sloan for electrostatic
and weakly electromagnetic instabilities \cite{1.1693669}. It will
be confirmed quantitatively in the context of the electromagnetic
ITG instability below.

It is worth noticing that a similar stabilization can be expected
when a population of fast ions is present \cite{0741-3335-52-4-045007}.
Just like the electrons, fast ions move quickly along the magnetic
field, and in addition they have large gyroradii. If the typical velocity
of the fast ions exceeds the phase velocity of the instability along
the field, $\omega/k_{\|}v_{fast}<1$, or their gyroradius exceeds
the perpendicular wavelength, $k_{\perp}\rho_{fast}>1$, such ions
will experience relatively small ${\bf E}\times{\bf B}$ displacement
and therefore contribute little to the magnetic-drift perturbation
in Eq.~\eqref{eq:vdriftpert}. The fast ions will then contribute
relatively little to the instability. On the other hand, their equilibrium
pressure can be significant and acts to reduce the \textit{equilibrium
}drift in Eq. \eqref{eq:vdrift} by a new additive term giving $p_{i}+p_{e}\rightarrow p_{i}+p_{e}+p_{fast}$.
We thus expect a net stabilising action from fast ions. Gyrokinetic
simulations of plasmas with such particles indeed indicate the presence
of a critical $\beta$ for electromagnetic ITG stabilization that
decreases with $L_{p}/R$ \cite{PhysRevLett.111.155001}. Moreover,
the effect of fast ions is even more significant in nonlinear simulations.

\section{Reduction of the gyrokinetic equations}

Bearing in mind the qualitative picture from the preceding Section,
we now give quantitative substance to our findings. We proceed by
first deriving from gyrokinetics a set of second order differential
equations for the electrostatic and the magnetic potentials. These
equations support the electrostatic ITG mode in the limit of vanishing
$\beta,$ Alfvénic perturbations, magnetic compessibility and finite-ion-Larmor
radius effects. They are derived in a large-$\eta_{i}$ expansion,
where $\eta_{i}=d\log T_{i}/d\log n_{i}=L_{n_{i}}/L_{T_{i}},$ with
$T_{i}$ and $n_{i}$ the equilibrium temperature and density, respectively.
Kinetic ballooning modes are therefore diamagnetically stabilized
within our ordering.

Our starting point is the linearized gyrokinetic equation in ballooning
space \cite{1.863121,0032-1028-23-7-005,0029-5515-20-11-011}

\begin{equation}
\begin{split} & iv_{\parallel}\nabla_{\parallel}h_{s}+\left(\omega-\hat{\omega}_{ds}\right)h_{s}=\left(\omega-\omega_{*s}^{T}\right)\frac{eF_{0s}}{T_{0s}}\times\\
 & \left\{ J_{0}\left(a_{s}\right)\left(\phi-v_{\parallel}A_{\parallel}\right)+\frac{T_{s}}{e}2\frac{v_{\perp}^{2}}{v_{ths}^{2}}\frac{J_{1}\left(a_{s}\right)}{a_{s}}\frac{\delta B_{\parallel}}{B}\right\} ,
\end{split}
\label{eq:GKions}
\end{equation}
where $\phi$ is the electrostatic potential, $A_{\parallel}$ the
perturbed magnetic potential parallel to the equilibrium magnetic
field in the Coulomb gauge, $\nabla\cdot\mathbf{A}=0$, $\delta B_{\parallel}$
the parallel magnetic field perturbation, and $B$ the modulus of
the equilibrium magnetic field. The form of the perturbations used
is $\sim\exp[-i\omega t+i\mathbf{k}_{\perp}\cdot\mathbf{x}].$ The
function $h_{s},$ defined by $h_{s}\exp\left(-iL_{s}\right)=\delta f_{s}+Z_{s}e\phi F_{0s}/T_{s},$
denotes the nonadiabatic part of the perturbed distribution function,
$\delta f_{s}$, where $f_{s}=F_{0s}+\delta f_{s},$ with $\delta f_{s}\ll F_{0s},$
$F_{0s}$ is a Maxwellian equilibrium with temperature $T_{s}=m_{s}v_{ths}^{2}/2$
and density $n_{0s}$, $L_{s}=\mathbf{k}\times\mathbf{v}_{\perp}\cdot\hat{\mathbf{b}}/\Omega_{s},$
$\hat{\mathbf{b}}=\mathbf{B}/B,$ with $\mathbf{v}_{\perp}$ the perpendicular
particle velocity. Here $\Omega_{s}=Z_{s}eB/m_{s}$ is the cyclotron
frequency, $J_{0}$ and $J_{1}$ are Bessel function of the first
kind of argument $a_{s}=\hat{v}_{\perp}k_{\perp}\rho_{s}\equiv\hat{v}_{\perp}\sqrt{2b},$
where $\rho_{s}=v_{ths}/\Omega_{s}$ is the Larmor radius, $k_{\perp}^{2}=k_{y}^{2}(1+\hat{s}^{2}z^{2}),$
with $k_{y}$ the mode wave number, $\hat{s}$ the local magnetic
shear and $z$ the distance along the equilibrium field lines. Furthermore,
$\hat{\omega}_{ds}=2\left(\omega_{B}\hat{v}_{\perp}^{2}/2+\omega_{\kappa}\hat{v}_{\parallel}^{2}\right),$
$2\omega_{B}=\mathbf{k}_{\perp}\rho_{s}\cdot v_{ths}\hat{\mathbf{b}}\times\nabla B/B,$
$2\omega_{\kappa}=\mathbf{k}_{\perp}\rho_{s}\cdot v_{ths}\hat{\mathbf{b}}\times\left(\hat{\mathbf{b}}\cdot\nabla\hat{\mathbf{b}}\right),$
with $v_{\parallel}$ the parallel particle velocity. Finally, $\omega_{*s}^{T}=\omega_{*s}+\eta_{s}\omega_{*s}\left(\hat{v}^{2}-3/2\right),$
$\hat{v}=v/v_{ths},$ and $\omega_{*s}=(1/2)k_{y}\rho_{s}v_{ths}/L_{n_{s}}.$

The gyrokinetic equation \eqref{eq:GKions} is most easily solved
for the electrons, which we take to be sufficiently light that the
terms multiplied by $v_{\|}$ dominate. Neglecting magnetic trapping,
we thus obtain the electron response being described by the solution
\begin{equation}
h_{e}\approx-\left(1-\frac{\omega_{*e}^{T}}{\omega}\right)\frac{e\psi}{T_{e}}F_{0e},\label{eq:elsol}
\end{equation}
where we have written $\nabla_{\parallel}\psi=i\omega A_{\parallel}$.

For the ions, Eq. \eqref{eq:GKions} is solved iteratively using the
ordering \cite{1.859023,1.4868412} 
\begin{equation}
\frac{k_{\parallel}^{2}v_{thi}^{2}}{\omega^{2}}\sim\frac{\omega}{\eta_{i}\omega_{*i}}\sim\frac{\omega_{\kappa}+\omega_{B}}{\omega}\sim b\sim\epsilon\ll1,\label{eq:ordering}
\end{equation}
which retains the strongly driven ($\eta_{i}\gg1$) toroidal and slab
ITG instability and finite Larmor radius (FLR) effects. To include
electromagnetic perturbations in the electrostatic picture, we use
a maximal ordering for the fields, $v_{thi}A_{\parallel}\sim\epsilon\phi$,
and find in lowest order 
\begin{equation}
h_{i}^{(0)}=\frac{\omega-\omega_{*i}^{T}}{\omega-\hat{\omega}_{di}}\left[J_{0}\left(a_{i}\right)\frac{e\phi}{T_{0i}}+2\hat{v}_{\perp}^{2}\frac{J_{1}\left(a_{i}\right)}{a_{i}}\frac{\delta B_{\parallel}}{B}\right]F_{0i}.\label{eq:ionzero}
\end{equation}
The electrostatic potential is obtained from the quasineutrality condition,
\begin{equation}
n_{0}e(T_{e}^{-1}+T_{i}^{-1})\phi=\int d^{3}vJ_{0}h_{i}-\int d^{3}vh_{e},\label{eq:qn}
\end{equation}
to which the contribution from $h_{i}^{(0)}$ becomes 
\[
\int d^{3}vJ_{0}h_{i}^{(0)}=\frac{e\phi}{T_{i}}\left[1-\frac{\omega_{*i}}{\omega}(1-\eta_{i}b)-\frac{\eta_{i}\omega_{*i}(\omega_{B}+\omega_{\kappa})}{\omega^{2}}\right]-\frac{\eta_{i}\omega_{*i}}{\omega}\frac{\delta B_{\parallel}}{B}.
\]
in lowest order. This density perturbation is a factor $\eta_{i}^{-1}\ll1$
smaller than expected from the size of $h_{i}^{(0)}\sim(\eta_{i}\omega_{*i}/\omega)(e\phi/T_{i})F_{0i}$,
compelling us to find the solution to higher order. We thus iterate
the solution, 
\begin{equation}
\begin{split} & h_{i}\approx h_{i}^{(0)}-\frac{v_{\parallel}}{\omega-\hat{\omega}_{di}}\left\{ \left(\omega-\omega_{*i}^{T}\right)J_{0}\left(a_{i}\right)\frac{eA_{\parallel}}{T_{i}}F_{0i}+i\nabla_{\parallel}\biggl[h_{i}^{(0)}\,\,\,\right.\\
 & \left.\left.-\frac{v_{\parallel}}{\omega-\hat{\omega}_{di}}\left(\left(\omega-\omega_{*i}^{T}\right)J_{0}\left(a_{i}\right)\frac{eA_{\parallel}}{T_{i}}F_{0i}+i\nabla_{\parallel}h_{i}^{(0)}\right)\right]\right\} ,
\end{split}
\label{eq:ionsol}
\end{equation}
and find that a sufficiently accurate expression for the ion density
perturbation is 
\[
\int d^{3}vJ_{0}h_{i}=\frac{e\phi}{T_{i}}\left[1-\frac{\omega_{*i}}{\omega}(1-\eta_{i}b)-\frac{\eta_{i}\omega_{*i}(\omega_{B}+\omega_{\kappa})}{\omega^{2}}\right]
\]
\begin{equation}
-\frac{\eta_{i}\omega_{*i}}{\omega}\frac{\delta B_{\parallel}}{B}+\frac{\eta_{i}\omega_{*i}B}{m_{i}\omega^{3}}\nabla_{\|}\left[\frac{e\nabla_{\|}(\phi-\psi)}{B}\right],\label{eq:ion density}
\end{equation}

The magnetic field strength fluctuations are determined by the perpendicular
Amp\`ere's law, 
\begin{equation}
\frac{\delta B_{\parallel}}{B}=-\frac{\mu_{0}}{B^{2}}\sum_{s}\int d^{3}vm_{s}v_{\perp}^{2}a_{s}^{-1}J_{1}\left(a_{s}\right)h_{s},
\end{equation}
implying that $\delta B_{\parallel}/B$ is proportional to $\beta e\phi/T_{i}.$
At this point, a traditional and popular approach would be to neglect
the magnetic compressibility altogether \cite{1.860623,1.1730294,0741-3335-56-1-015007,0741-3335-55-12-125003,1.1342029,1.4876960},
since $\beta$ in fusion relevant plasmas is of the order of $1\%-5\%.$
However, even such a small $\beta$ is not necessarily negligible,
since it gets multiplied by a large factor of order $\epsilon^{-1}$
in Eq.~\eqref{eq:ion density}. In fact, using Eqs. \eqref{eq:elsol}
and \eqref{eq:ionsol} to calculate the integrals in Amp\`ere's law,
we arrive at the conclusion that $\beta\sim\omega^{2}/(\eta_{i}^{2}\omega_{*i}^{2})\sim\epsilon^{2}\ll1$
is the correct ordering that allows us to calculate 
\begin{equation}
\frac{T_{i}}{e}\frac{\delta B_{\parallel}^{(1)}}{B}=\frac{\beta_{i}}{2}\frac{\eta_{i}\omega_{*i}}{\omega}\left(\phi+\frac{1}{\tau}\frac{\eta_{e}}{\eta_{i}}\psi\right),\label{eq:delB1}
\end{equation}
where $\beta_{i}=2\mu_{0}nT_{i}/B^{2}$ and $\tau=T_{i}/T_{e}$. This
result is a special case of a general formula derived in the work
of Tang et al. \cite{0029-5515-20-11-011} on kinetic ballooning modes.
Finally, using Eqs. \eqref{eq:elsol}, \eqref{eq:ion density} and
\eqref{eq:delB1} in the quasineutrality condition \eqref{eq:qn},
we obtain 
\begin{equation}
\begin{split}\left[\tau+\frac{\omega_{*i}}{\omega}-\frac{\beta_{i}}{2\tau}\frac{\eta_{e}}{\eta_{i}}\frac{\eta_{i}^{2}\omega_{*i}^{2}}{\omega^{2}}-\frac{\eta_{i}\omega_{*i}v_{thi}^{2}}{2\omega^{3}l_{\parallel}^{2}}\frac{\partial^{2}}{\partial z^{2}}\right]\left(\phi-\psi\right)=-\left(2\eta_{i}\frac{\omega_{*i}\omega_{\kappa}}{\omega^{2}}-\eta_{i}\frac{\omega_{*i}}{\omega}b\right)\phi\end{split}
\label{eq:QNbef-1}
\end{equation}
where we have defined the normalization length $l_{\parallel}$ and
the coordinate $z$ along the field so that $l_{\parallel}\nabla_{\parallel}\equiv\partial_{z}$.
We have also used the result 
\begin{equation}
\omega_{\kappa}-\omega_{B}=[1+\eta_{e}/(\tau\eta_{i})]\eta_{i}\omega_{*i}\beta_{i}/2,\label{eq:relazione}
\end{equation}
which follows directly from the force balance equation \cite{0029-5515-20-11-011}
\begin{equation}
\bm{j}\times\bm{B}=\nabla p.
\end{equation}
Equation \eqref{eq:QNbef-1} is similar to previous results in the
literature, but is different in a couple of ways. In particular, the
third term on the LHS is absent from previous electromagnetic theories
of ITG instabilities \cite{1.860623}. Another novelty of this equation
is that the inclusion of the ion contribution to magnetic compressibility
{[}the term proportional to $\phi$ in Eq. \eqref{eq:delB1}{]} resulted
in the ``rule'' that the drive of the toroidal branch of the ITG
{[}the first term on the RHS of Eq. \eqref{eq:QNbef-1}{]} is the
curvature drift only. This result has been confirmed by various numerical
works \cite{1.873694,1.3432117,Tobias}.

We close the system of equations calculating the divergence of the
current \cite{1.863121,0029-5515-20-11-011} to obtain 
\begin{equation}
\begin{split}\frac{1}{\beta_{i}B}\frac{v_{thi}^{2}/l_{\parallel}^{2}}{\omega^{2}}\frac{\partial}{\partial z}\left(bB\frac{\partial}{\partial z}\psi\right)=b\eta_{i}\frac{\omega_{*i}}{\omega}\phi-2\frac{\eta_{i}\omega_{*i}\omega_{\kappa}}{\omega^{2}}\left(\phi+\frac{1}{\tau}\frac{\eta_{e}}{\eta_{i}}\psi\right).\end{split}
\label{eq:currdivbef-1}
\end{equation}
This is obtained by using Amp\`ere's law after taking the $\Sigma_{s}e_{s}\int d^{3}vJ_{0}$
moment of the gyrokinetic equation \eqref{eq:GKions} 
\begin{equation}
\begin{split} & \frac{B}{\mu_{0}\omega^{2}}\nabla_{\text{\ensuremath{\parallel}}}\left(\frac{k_{\perp}^{2}}{B}\nabla_{\text{\ensuremath{\parallel}}}\psi\right)=\frac{ne^{2}}{T_{i}}\eta_{i}\frac{\omega_{*i}}{\omega}b\phi+\sum_{s}e_{s}\int d^{3}vJ_{0}\frac{J_{1}\left(a_{i}\right)}{a_{i}}\frac{\omega_{*s}^{T}}{\omega}2\hat{v}_{\perp}^{2}\frac{\delta B_{\parallel}}{B}+\sum_{s}e_{s}\int d^{3}vJ_{0}\frac{\hat{\omega}_{ds}}{\omega}h_{s},\end{split}
\label{eq:divgen}
\end{equation}
where the ordering in Eq. \eqref{eq:ordering} as been used. Velocity-space
integrals are performed using solutions \eqref{eq:elsol} and \eqref{eq:ionzero}.
Thus, from Eq. \eqref{eq:divgen}, we have 
\begin{equation}
\begin{split} & \frac{B}{\mu_{0}\omega^{2}}\nabla_{\text{\ensuremath{\parallel}}}\left(\frac{k_{\perp}^{2}}{B}\nabla_{\text{\ensuremath{\parallel}}}\psi\right)=\eta_{i}\frac{\omega_{*i}}{\omega}b\phi-\left(1+\frac{1}{\tau}\frac{\eta_{e}}{\eta_{i}}\right)\eta_{i}\frac{\omega_{*i}}{\omega}\frac{T_{i}}{e}\frac{\delta B_{\parallel}^{(1)}}{B}\\
 & -\frac{\eta_{i}\omega_{*i}(\omega_{\kappa}+\omega_{B})}{\omega^{2}}\phi-\frac{1}{\tau}\frac{\eta_{e}}{\eta_{i}}\frac{\eta_{i}\omega_{*i}(\omega_{\kappa}+\omega_{B})}{\omega^{2}}\psi.
\end{split}
\end{equation}
Substitution of Eqs. \eqref{eq:delB1} and \eqref{eq:relazione} now
yields Eq. \eqref{eq:currdivbef-1}.

\section{Critical $\beta$ for stability\label{sec:shearless}}

Let us first consider the simple case in which the equilibrium magnetic
field is independent of the coordinate along $\bm{B}$. Then $\hat{s}\equiv0,$
$k_{\perp}^{2}=k_{y}^{2},$ and we can Fourier transform Eqs. \eqref{eq:QNbef-1}-\eqref{eq:currdivbef-1},
to obtain 
\begin{equation}
\left(\tau+\frac{\omega_{*i}}{\omega}-\frac{\beta_{i}}{2\tau}\frac{\eta_{e}}{\eta_{i}}\frac{\omega_{T}^{2}}{\omega^{2}}+\frac{\omega_{T}}{2\omega}\frac{k_{z}^{2}v_{thi}^{2}}{\omega^{2}}\right)\Lambda=-2\frac{\omega_{T}\omega_{\kappa}}{\omega^{2}}+b\frac{\omega_{T}}{\omega},\label{eq:genshearless}
\end{equation}
where $\omega_{T}=\eta_{i}\omega_{*i},$ 
\begin{equation}
\Lambda=\frac{\left(1+\frac{\eta_{e}}{\tau\eta_{i}}\right)\left(\beta_{{\scriptscriptstyle MHD}}-\beta_{i}\right)+\frac{\omega}{2\omega_{\kappa}}b\beta_{i}}{\frac{\eta_{e}}{\tau\eta_{i}}\left(\beta_{{\scriptscriptstyle MHD}}-\beta_{i}\right)+\beta_{{\scriptscriptstyle MHD}}},
\end{equation}
and 
\begin{equation}
\beta_{{\scriptscriptstyle MHD}}=b\frac{k_{z}^{2}v_{thi}^{2}}{2\omega_{\kappa}\eta_{i}\omega_{*i}[1+\eta_{e}/(\tau\eta_{i})]}\label{eq:betaMHD}
\end{equation}
is the value of $\beta$ above which ideal MHD modes would be destabilized
if they were not suppressed by diamagnetic effects.

For $\beta_{{\scriptscriptstyle MHD}}\ll\beta_{i},$ $\Lambda\approx\tau\eta_{i}[1+\eta_{e}/(\tau\eta_{i})-b\omega/(2\omega_{\kappa})]/\eta_{e}$,
while for $\beta_{{\scriptscriptstyle MHD}}\gg\beta_{i},$ $\Lambda\approx1.$
Similarly, for $\beta_{i}\rightarrow0,$ $\Lambda\rightarrow1,$ and
Eq. \eqref{eq:genshearless} reduces to the dispersion relation for
the electrostatic ITG mode \cite{1.859023,1.1762151,1.4868412}, 
\begin{equation}
\tau+\frac{\omega_{*i}}{\omega}+\frac{\omega_{T}}{2\omega}\frac{k_{z}^{2}v_{thi}^{2}}{\omega^{2}}=-2\frac{\omega_{T}\omega_{\kappa}}{\omega^{2}}+b\frac{\omega_{T}}{\omega}.
\end{equation}

\begin{figure}[!h]
\scalebox{1.}{\input{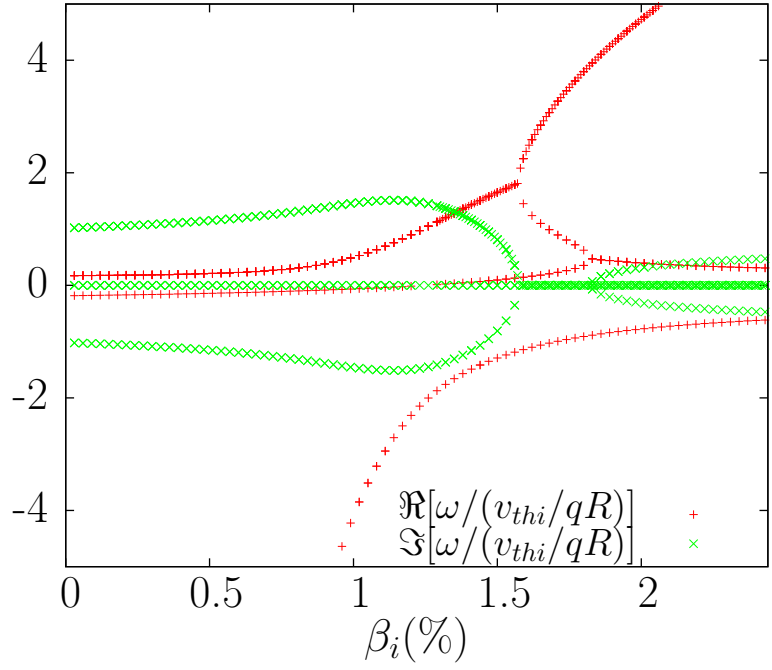}}

\caption{Real and imaginary part of the roots of Eq. \eqref{eq:genshearless}
for $\tau=1,$ $k_{z}l_{\parallel}\equiv k_{z}qR=0.5,$ $b=0.1,$
$R/L_{T_{i}}=5,$ $\omega_{T}/(v_{thi}/qR)=\sqrt{b}R/L_{T_{i}},$
$\omega_{\kappa}/(v_{thi}/qR)=\sqrt{b},$ $\omega_{*i}\eta_{e}=\omega_{T}.$
The unstable root at low-$\beta_{i}$ is the toroidal branch of the
ITG mode. The mode is stabilized for $\beta_{i}\approx1.5\%.$}

\label{Flo:2-1} 
\end{figure}

Equation \eqref{eq:genshearless} agrees with the large-$\eta_{i}$
limit of Eq. (25) in Ref. \cite{1.860623} only if the electron contribution
to the magnetic compressibility (the third term on the LHS) is neglected.
In general, the coupling of all the roots of Eq. \eqref{eq:genshearless}
is essential to understand the electromagnetic stabilization of the
toroidal ITG mode. To illustrate a somewhat typical case, we solve
Eq. \eqref{eq:genshearless} numerically for $\tau=1,$ $b=0.1,$
$R/L_{T_{i}}=5,$ $k_{z}l_{\parallel}\equiv k_{z}qR=0.5,$ and $q=\sqrt{2}$,
where $R$ is the major radius of the toroidal device, and $q$ measures
the pitch of the magnetic field. We consider the flat density limit
for simplicity, $\omega_{*i}\equiv0,$ but $\omega_{T}\neq0$. For
these values $\beta_{MHD}=0.0125.$ We note the normalised frequencies
$\omega_{T}/(v_{thi}/qR)=\sqrt{b/2}qR/L_{T_{i}}=\sqrt{b}R/L_{T_{i}}$
and $\omega_{\kappa}/(v_{thi}/qR)=q\sqrt{b/2}=\sqrt{b},$ for this
particular value of $q.$ Several electromagnetic branches can be
observed, depending on the value of $\beta_{i}.$ For small $\beta_{i}$
we find two complex conjugated ion roots. In Fig. \eqref{Flo:2-1}
we identify the toroidal ITG branch, $0<\Re[\omega/(v_{thi}/qR)]\ll1,$
and $0<\Im[\omega/(v_{thi}/qR)]<1.$ Its $\beta-$stabilization occurs
at a critical $\beta_{i}^{crit}$ for which the imaginary parts of
the two complex conjugated roots coalesce. At low $\beta_{i},$ a
further stable electron mode $\Re[\omega/(v_{thi}/qR)]<0$ is present.
Its real part changes sign when the $\beta-$stabilization of the
ITG becomes effective for $\beta_{i}\approx1\%$, see Fig. \eqref{Flo:2-1}.

\begin{figure}[!h]
\scalebox{1.}{\input{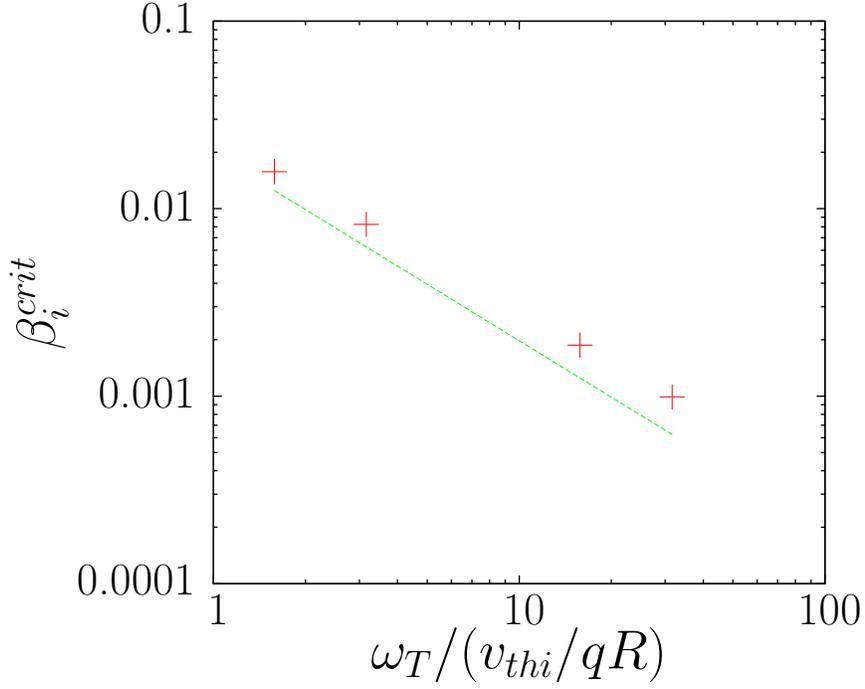}}

\caption{The critical $\beta_{i}$ for stabilization, as a function of $\omega_{T}/(v_{thi}/qR)=\sqrt{b/2}qR/L_{T_{i}},$
calculated from the solution of Eq. \eqref{eq:genshearless}. All
parameters are as in Fig. \eqref{Flo:2-1}. The line is from Eq. \eqref{eq:betaMHD}.}

\label{Flo:6-1} 
\end{figure}

\begin{figure}[!h]
\scalebox{1.}{\input{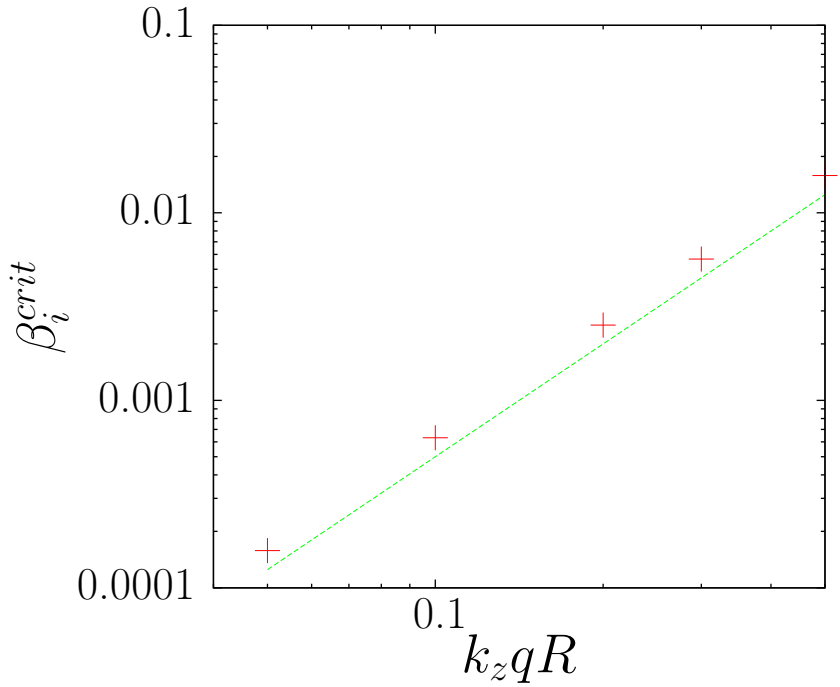}}

\caption{The critical $\beta_{i}$ for stabilization, as a function of $k_{z}l_{\parallel}\equiv k_{z}qR,$
calculated from the solution of Eq. \eqref{eq:genshearless}. All
parameters are as in Fig. \eqref{Flo:2-1}. The line is from Eq. \eqref{eq:betaMHD}.}

\label{Flo:7-1} 
\end{figure}

To establish the scaling of the observed $\beta$ for stabilization
with $\omega_{T}$, we solve Eq. \eqref{eq:genshearless} for several
values of $\omega_{T}/(v_{thi}/qR)=\sqrt{b/2}qR/L_{T_{i}}$ but fixing
$\omega_{\kappa}/(v_{thi}/qR)=q\sqrt{b/2}=\sqrt{.1}$, for the above
values of $b$ and $q.$ This means $\omega_{T}/(v_{thi}/qR)=\sqrt{0.1}R/L_{T_{i}}.$
We then record the value of $\beta_{i}$ at which the mode is completely
stable. To determine the scaling of the observed $\beta$ for stabilization
with $k_{z},$ we repeat the same evaluation of $\beta_{i}^{crit}$
for constant $\omega_{T}/(v_{thi}/qR)=\sqrt{b/2}qR/L_{T_{i}}=\sqrt{0.1}5,$
but varying $k_{z}Rq.$ As Figs. \eqref{Flo:6-1}-\eqref{Flo:7-1}
show, the critical $\beta$ for stabilization scales as $\beta_{i}^{crit}\sim\beta_{{\scriptscriptstyle MHD}}$,
which implies \cite{wesson} 
\begin{equation}
\beta_{i}^{crit}\sim\frac{1}{2q^{2}}\frac{L_{T_{i}}}{R}.\label{eq:betacritions}
\end{equation}
However, as is evident from the figures, $\beta_{i}^{crit}$ lies
somewhat above $\beta_{{\scriptscriptstyle MHD}},$ which means that,
for these parameters, the stabilization occurs only for values of
$\beta_{i}$ above the ideal MHD threshold. 

It is interesting to analyze the stability below this threshold, for
$\beta_{i}\ll\beta_{{\scriptscriptstyle MHD}}.$ This situation corresponds
to $\Lambda\approx1.$ For a strongly toroidal mode $4\omega_{T}\omega_{\kappa}\gg k_{z}^{2}v_{thi}^{2}\omega_{T}/\omega,$
or 
\begin{equation}
\beta_{i}\ll\beta_{{\scriptscriptstyle MHD}}\ll b\frac{\omega}{\omega_{T}}\approx b\sqrt{\frac{\omega_{\kappa}}{\omega_{T}}},
\end{equation}
the new term on the LHS of Eq. \eqref{eq:genshearless} cannot be
neglected, and indeed it is responsible for a new critical \textsl{electron}
$\beta_{e}$ for stabilization. After neglecting the stabilizing FLR
term on the RHS of Eq. \eqref{eq:genshearless}, we obtain $\tau\omega^{2}=-2\omega_{T}\omega_{\kappa}/\Lambda+\beta_{i}\eta_{e}\omega_{T}^{2}/(2\tau\eta_{i}).$
Hence, the electron contribution to magnetic compressibility suppresses
the instability when 
\begin{equation}
\beta_{e}>\beta_{e}^{crit}=\frac{\eta_{i}}{\eta_{e}}\frac{4\omega_{\kappa}}{\Lambda\omega_{T}},\,\,\mbox{for}\,\Lambda>0.\label{eq:crit}
\end{equation}
In the limit $\beta_{i}\ll\beta_{{\scriptscriptstyle MHD}},$ $\Lambda\approx1.$
The same critical $\beta_{e}^{crit}$ for stabilization is obtained
in the $\beta_{{\scriptscriptstyle MHD}}\ll\beta_{i}$ limit, but
now $\Lambda\neq1.$ In both cases, we find 
\begin{equation}
\beta_{e}^{crit}\sim\frac{L_{T_{e}}}{R}.\label{eq:betacritelectrons}
\end{equation}
It is perhaps interesting to notice that $\beta_{{\scriptscriptstyle i}}^{crit}$
and $\beta_{{\scriptscriptstyle e}}^{crit}$ show different explicit
scalings with $\omega_{\kappa},$ however they follow the same scaling
with $R/L_{T}.$

\begin{figure}[!h]
\scalebox{1.}{\input{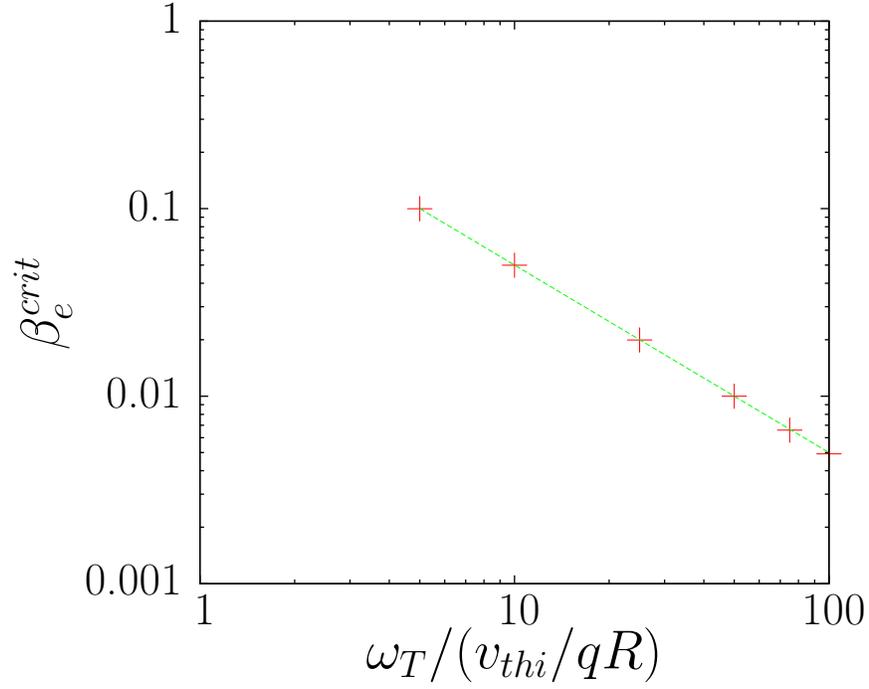}}

\caption{The critical $\beta_{i}$ for stabilization, as a function of $\omega_{T}/(v_{thi}/qR),$
calculated from the solution of Eq. \eqref{eq:genshearless}. Here
$k_{z}l_{\parallel}\equiv k_{z}qR=0.001,$ $\omega_{\kappa}/(v_{thi}/qR)=0.25,$
$\tau=1,$ $b=0.05,$ and $q=1.58.$ The line is from Eq. \eqref{eq:crit}
with $\Lambda=2.$}

\label{Flo:6} 
\end{figure}

\begin{figure}[!h]
\scalebox{1.}{\input{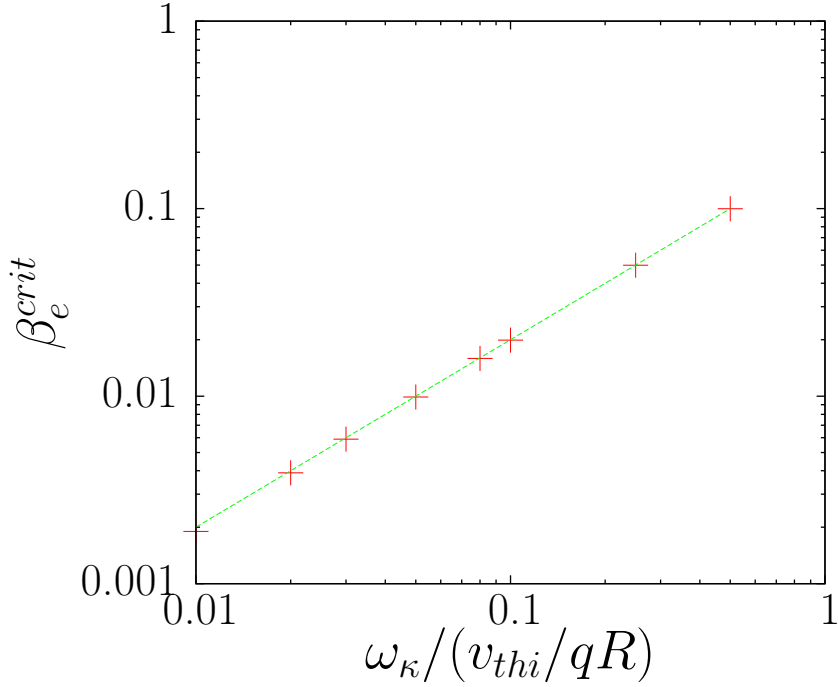}}

\caption{The critical $\beta_{i}$ for stabilization, as a function of $\omega_{\kappa}/(v_{thi}/qR)=q\sqrt{b/2},$
calculated from the solution of Eq. \eqref{eq:genshearless}. Here
$k_{z}qR=0.001,$ $\omega_{T}/(v_{thi}/qR)=10,$ $\tau=1,$ and $q=1.58.$
The line is from Eq. \eqref{eq:crit} with $\Lambda=2.$}

\label{Flo:7} 
\end{figure}

To verify the estimate in Eq. \eqref{eq:crit}, we now solve Eq. \eqref{eq:genshearless}
numerically in the asymptotic regime $\omega_{*i}\equiv0,$ $k_{z}qR=0.001,$
$\omega_{T}/(v_{thi}/qR)=10,$ $\omega_{\kappa}/(v_{thi}/qR)=0.25,$
$\tau=1,$ $b=0.05,$ and $q=1.58.$ For these values $\Lambda=2-\omega/(2v_{thi}/qR)\approx2,$
when $\omega/(v_{thi}/qR)\ll1$. Again, we solve Eq. \eqref{eq:genshearless}
for several values of $\omega_{T}/(v_{thi}/qR)$ at fixed $\omega_{\kappa}/(v_{thi}/qR)=0.25$,
and $\omega_{\kappa}/(v_{thi}/qR)=q\sqrt{b/2}$ at fixed $\omega_{T}/(v_{thi}/qR)=10$,
and record the value of $\beta_{i}$ at which the mode is completely
stable. As Figs. \eqref{Flo:6}-\eqref{Flo:7} show, the critical
$\beta_{e}$ for stabilization agrees with Eq. \eqref{eq:crit}. In
Fig. \eqref{Flo:2-1}, we notice a window of stability for $\beta\approx1\%.$
A further destabilization might occur for $\beta>2\%.$ Incidentally,
the high-$\beta$ unstable mode is \textsl{not }the kinetic balloning
mode, since this is diamagnetically stabilized in our large-$\eta_{i}$
limit. The presence of electromagnetic roots can be investigated further
by considering the limit $\beta_{i}\rightarrow\beta_{MHD}.$ 

\subsection{Electromagnetic roots at $\beta_{i}\approx\beta_{{\scriptscriptstyle MHD}}$}

Equation \eqref{eq:genshearless}, when $\omega_{*i}\equiv0,$ is
in general a quartic for $\omega.$ However, near $\beta_{i}\approx\beta_{{\scriptscriptstyle MHD}},$
it can be factored into a stable solution
\begin{equation}
\omega_{stab}=-\frac{k_{z}^{2}v_{thi}^{2}}{4\omega_{\kappa}^{2}}\left(1+\frac{\eta_{e}}{\tau\eta_{i}}\right)\frac{\beta_{{\scriptscriptstyle MHD}}-\beta_{i}}{1+b\frac{k_{z}^{2}v_{thi}^{2}}{8\omega_{\kappa}^{2}}},
\end{equation}
and a cubic 
\begin{equation}
\omega^{3}+a_{1}\omega+a_{0}=0,\label{eq:cubic}
\end{equation}
with 
\begin{equation}
a_{1}=-\frac{1}{\tau}\left(\frac{1}{1+\tau\eta_{i}/\eta_{e}}\frac{k_{z}^{2}v_{thi}^{2}}{4\omega_{\kappa}^{2}}b+2\right)\omega_{\kappa}\omega_{T},
\end{equation}
and 
\begin{equation}
a_{0}=\frac{1}{\tau}\left(\frac{\omega_{T}k_{z}^{2}v_{thi}^{2}}{2}+4\frac{\omega_{\kappa}^{2}\omega_{T}}{b}\right).
\end{equation}

For $\beta_{i}\rightarrow\beta_{{\scriptscriptstyle MHD}},$ $\omega_{stab}\rightarrow0,$
whereas the roots of Eq. \eqref{eq:cubic} are 
\begin{equation}
\omega_{1}=A+B,
\end{equation}
\begin{equation}
\omega_{2}=-\frac{1}{2}\left(A+B\right)+i\frac{\sqrt{3}}{2}\left(A-B\right),
\end{equation}
and 
\begin{equation}
\omega_{3}=-\frac{1}{2}\left(A+B\right)-i\frac{\sqrt{3}}{2}\left(A-B\right),
\end{equation}
with $A^{3}=C+\sqrt{C^{2}+D^{3}},$ $B^{3}=C-\sqrt{C^{2}+D^{3}},$
\begin{equation}
C=-\frac{1}{2}a_{0}<0,
\end{equation}
and 
\begin{equation}
D=\frac{1}{3}a_{1}<0.
\end{equation}

For 
\begin{equation}
\left(\frac{a_{1}}{3}\right)^{3}+\left(\frac{a_{0}}{2}\right)^{2}<0,
\end{equation}
all three roots are real.

In the limit $(a_{1}/3)^{3}\ll(a_{0}/2)^{2},$ $A^{3}\sim a_{0}(a_{1}^{3}/a_{0}^{2})\ll1$,
and $B^{3}\approx-a_{0}=\mathcal{O}(1).$ Thus, we find the unstable
mode 
\begin{equation}
\omega_{1}\approx e^{i\frac{\pi}{3}}\left[\frac{1}{\tau}\left(\frac{\omega_{T}k_{z}^{2}v_{thi}^{2}}{2}+4\frac{\omega_{\kappa}^{2}\omega_{T}}{b}\right)\right]^{1/3},
\end{equation}
while $\omega_{2}=\omega_{1}^{*}$ is damped, and $\omega_{3}=-\left|\omega_{1}\right|$
is marginally stable. In the case of negligible slab drive, we have
\begin{equation}
\begin{split} & \omega_{1}\approx e^{i\frac{\pi}{3}}\left(\frac{4}{\tau}\frac{\omega_{\kappa}^{2}\omega_{T}}{b}\right)^{1/3}\approx e^{i\frac{\pi}{3}}\left(\frac{k_{y}\rho_{i}}{\tau}\right)^{1/3}\frac{v_{thi}}{R^{2/3}L_{T_{i}}^{1/3}},\,\mbox{for}\,\,\frac{R}{L_{T_{i}}}\frac{b^{2}}{\tau}\ll\frac{27}{2}\end{split}
.
\end{equation}

In the opposite limit $(a_{1}/3)^{3}\gg(a_{0}/2)^{2},$ we have $A^{3}\approx-i(\left|a_{1}\right|/3)^{3/2},$
and $B^{3}\approx-A^{3}.$ Therefore, we obtain one stable ion root
\begin{equation}
\omega_{1}\approx\left[\frac{1}{\tau}\left(\frac{1}{1+\tau\eta_{i}/\eta_{e}}\frac{k_{z}^{2}v_{thi}^{2}}{4\omega_{\kappa}^{2}}b+2\right)\omega_{\kappa}\omega_{T}\right]^{1/2},
\end{equation}
which, for negligible slab drive is 
\begin{equation}
\begin{split} & \omega_{1}\approx\sqrt{\frac{1}{\tau}\omega_{\kappa}\omega_{T}}\approx\frac{k_{y}\rho_{i}}{2}\frac{v_{thi}}{\sqrt{RL_{T_{i}}}},\,\mbox{for}\,\,\frac{R}{L_{T_{i}}}\frac{b^{2}}{\tau}\gg\frac{27}{2}.\end{split}
\end{equation}
The second root is 
\begin{equation}
\omega_{2}\approx0.
\end{equation}
Finally, we find the stable electron mode 
\begin{equation}
\omega_{3}=\text{-\ensuremath{\omega}}_{1}.
\end{equation}

\section{\textup{Electromagnetic boundary of marginal stability.} }

\begin{figure}
\includegraphics[scale=1.1]{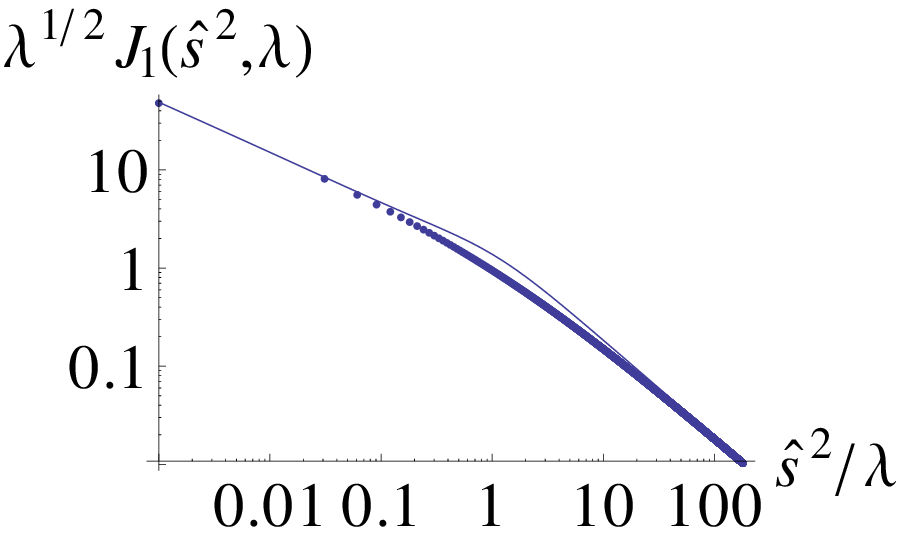}\caption{Comparison of the numerical solution (points) and analytical solution
(line) of $J_{1}$ as defined in Eq. \eqref{eq:jipade}.}

\label{fig:J1numvsan} 
\end{figure}

For the more realistic case of finite shear, we have $k_{\perp}^{2}=k_{y}^{2}\left(1+\hat{s}^{2}z^{2}\right),$
and the previous analysis does not apply. Nevertheless, we can still
construct a perturbative electromagnetic theory of the ITG instability
similar to that introduced in Ref. \cite{0741-3335-55-12-125003},
if we use a local approximation of the curvature drift, $\omega_{\kappa}(z)=\omega_{\kappa}(1-az^{2})$
\cite{1.4868412}. We calculate the electromagnetic correction to
the electrostatic eigenvalue using a low $\beta_{i}\omega_{T}^{2}/\omega^{2}\ll1$
subsidiary expansion. The zeroth order electrostatic response is given
by Eq. \eqref{eq:QNbef-1} with $\psi^{(0)}=0,$ and $\phi^{(0)}=\exp[-\lambda z^{2}],$
with \cite{1.4868412} 
\begin{equation}
4\lambda^{2}=-2a\omega_{\kappa}\omega_{0}/\omega_{tr}^{2}-(b_{0}\omega_{0}^{2}/\omega_{tr}^{2})\hat{s}^{2},\label{eq:es1}
\end{equation}
\begin{equation}
\tau+\omega_{*i}/\omega_{0}+2\omega_{T}\omega_{\kappa}/\omega_{0}^{2}-b_{0}\omega_{T}/\omega_{0}+2\lambda\omega_{T}\omega_{tr}^{2}/\omega_{0}^{3}=0,\label{es2}
\end{equation}
and $\text{\ensuremath{\omega}}_{tr}^{2}=v_{tr}^{2}/(2l_{\parallel}^{2}).$
Equations \eqref{eq:es1}-\eqref{es2} constitute the electrostatic
eigenvalue equation, they determine $\lambda$ and $\omega_{0}$ which
have complex values. After writing Eq. \eqref{eq:QNbef-1} to first
order, we can calculate $\delta\omega$ so that $\omega=\omega_{0}+\delta\omega,$
with $\delta\omega/\omega_{0}=\mathcal{O}(\beta_{i}).$ Since the
zeroth order operator acting on $\phi^{(0)},$ $\mathcal{L}_{0}^{(0)}=-(\tau+\omega_{*i}/\omega_{0})-[2\omega_{T}\omega_{\kappa}(z)/\omega_{0}^{2}-b\omega_{T}/\omega_{0}]+\omega_{T}v_{thi}^{2}/(2l_{\parallel}^{2}\omega_{0}^{3})\partial_{z}^{2},$
is self-adjoint, we obtain 
\begin{equation}
\begin{split} & \frac{\delta\omega}{\omega_{0}}=-\left\{ \int_{-\infty}^{\infty}dz\phi^{(0)}\mathcal{L}_{1}^{(0)}\left[\phi^{(0)}\right]\right\} ^{-1}\times\\
 & \int_{-\infty}^{\infty}dz\phi^{(0)}\left\{ \left[\frac{\omega_{T}}{\omega_{0}}\frac{v_{thi}^{2}/l_{\parallel}^{2}}{2\omega_{0}^{2}}\frac{\partial^{2}}{\partial z^{2}}-\tau\left(1-\frac{\omega_{*e}}{\omega_{0}}\right)\right]\psi^{(1)}\left.-\frac{\beta_{i}}{2}\frac{\eta_{e}}{\tau\eta_{i}}\left(\frac{\omega_{T}}{\omega_{0}}\right)^{2}\phi^{(0)}\right\} ,\right.
\end{split}
\label{eq:emcorr}
\end{equation}
where 
\begin{equation}
\mathcal{L}_{1}^{(0)}=3\frac{\omega_{T}}{\omega_{0}}\frac{v_{thi}^{2}/l_{\parallel}^{2}}{2\omega_{0}^{2}}\frac{\partial^{2}}{\partial z^{2}}-\frac{\omega_{*i}}{\omega_{0}}-4\frac{\omega_{T}\omega_{\kappa}(z)}{\omega_{0}^{2}}+\frac{\omega_{T}}{\omega_{0}}b.
\end{equation}
Note that the expression for $\delta\omega$ only requires knowledge
of the eigenfunction $\phi^{(0)}$ to zeroth order. To perform the
integrations in Eq. \eqref{eq:emcorr}, we need the first order electromagnetic
component, $\psi^{(1)},$ given by Eq. \eqref{eq:currdivbef-1}. We
find 
\begin{equation}
\psi^{(1)}=\frac{\beta_{i}}{b_{0}}\frac{\omega_{0}^{2}}{v_{thi}^{2}/l_{\parallel}^{2}}\int_{0}^{z}dz\frac{\mu ze^{-\lambda z^{2}}+\nu\mathnormal{Erf}\left(\sqrt{\lambda}z\right)}{1+\hat{s}^{2}z^{2}},\label{eq:psi1}
\end{equation}
with $2\mu=-b_{0}\hat{s}^{2}\omega_{T}/(\omega_{0}\lambda)-2\omega_{\kappa}\omega_{T}a/(\lambda\omega_{0})^{2},$
and 
\begin{equation}
\nu=\frac{\sqrt{\pi}}{4\lambda}\,\left[b_{0}\frac{\omega_{T}}{\omega_{0}}\left(\hat{s}^{2}+2\lambda\right)+2\frac{\omega_{T}\omega_{\kappa}}{\omega_{0}^{2}}\left(a-2\lambda\right)\right].
\end{equation}
Thus, the electromagnetic correction to the electrostatic ITG for
finite shear is 
\begin{equation}
\begin{split} & \frac{\delta\omega}{\omega_{0}}=\sqrt{\frac{2}{\pi}}\left\{ \frac{\omega_{T}}{\omega_{0}}\lambda\frac{\beta_{i}}{b_{0}}\left[\mu J_{3}+\nu J_{2}\right]+\tau\left(1-\frac{\omega_{*e}}{\omega_{0}}\right)\times\right.\\
 & \left.\sqrt{\frac{\pi}{\lambda}}\frac{\beta_{i}}{b_{0}}\frac{\omega_{0}^{2}}{v_{thi}^{2}/l_{\parallel}^{2}}\left[\mu J_{2}+\nu J_{1}\right]-\frac{\beta_{i}}{2}\frac{\eta_{e}}{\tau\eta_{i}}\frac{\omega_{T}^{2}}{\omega_{0}^{2}}\sqrt{\frac{\pi}{2\lambda}}\right\} \times\\
 & \left\{ 3\frac{\omega_{T}}{\omega_{0}}\frac{v_{thi}^{2}/l_{\parallel}^{2}}{2\omega_{0}^{2}}\lambda^{1/2}+\right.\left(\frac{\omega_{*i}}{\omega_{0}}+4\frac{\omega_{T}\omega_{\kappa}}{\omega_{0}^{2}}-\frac{\omega_{T}}{\omega_{0}}b_{0}\right)\frac{1}{\lambda^{1/2}}\\
 & \left.-\frac{1}{4}\left(4\frac{\omega_{T}\omega}{\omega_{0}^{2}}a+\frac{\omega_{T}}{\omega_{0}}b_{0}\hat{s}^{2}\right)\frac{1}{\lambda^{3/2}}\right\} ^{-1},
\end{split}
\label{eq:deomegafin}
\end{equation}
with $J_{1}=\int_{0}^{\infty}dz\mathnormal{Erf}^{2}\left(\sqrt{\lambda}z\right)(1+\hat{s}^{2}z^{2})^{-1},$
$J_{2}=\int_{0}^{\infty}dzzErf\left(\sqrt{\lambda}z\right)e^{-\lambda z^{2}}(1+\hat{s}^{2}z^{2})^{-1},$
and $J_{3}=\int_{0}^{\infty}dzz^{2}e^{-2\lambda z^{2}}(1+\hat{s}^{2}z^{2})^{-1}.$
We find an analytic closed form of Eq. \eqref{eq:deomegafin} if we
introduce the Padé approximants for the two asymptotic limits $\hat{s}^{2}\gg\lambda$
and $\hat{s}^{2}\ll\lambda.$ For the integral $J_{1},$ we find

\begin{equation}
\lambda^{1/2}J_{1}(\hat{s}^{2},\lambda)\simeq\frac{\left(\frac{\pi}{2}\frac{\lambda^{1/2}}{\hat{s}}-\sqrt{\frac{2}{\pi}}\right)+\frac{4}{\sqrt{\pi}}\log\left(1+\sqrt{2}\right)\frac{\hat{s}}{\lambda^{1/2}}}{1+\left(\frac{\hat{s}}{\lambda^{1/2}}\right)^{3}},\label{eq:jipade}
\end{equation}
see Fig. \eqref{fig:J1numvsan}. 

\begin{figure}
\includegraphics[scale=1.1]{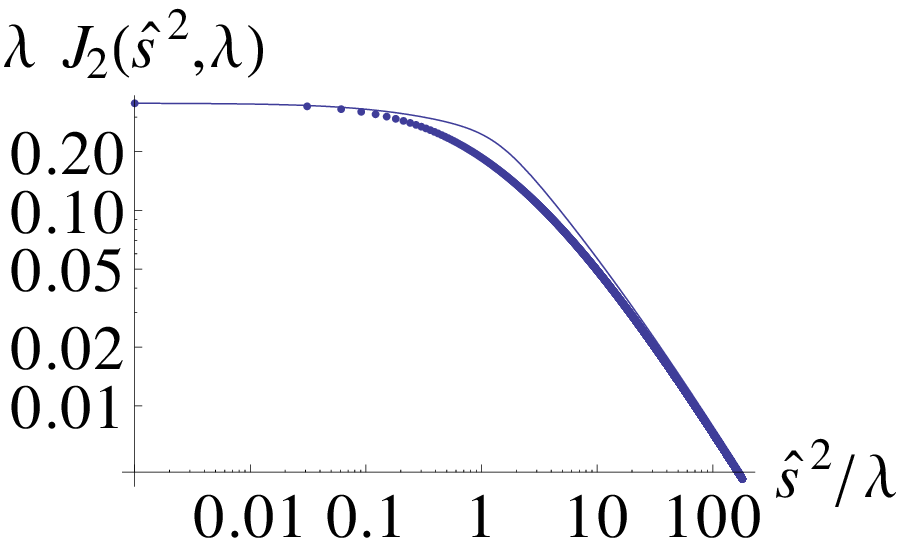}\caption{Comparison of the numerical solution (points) and the analytical solution
(line) of $J_{2}$ as defined in Eq. \eqref{eq:j2pade}.}

\label{fig:J2numvsan} 
\end{figure}

The Padé approximant of $J_{2,}$ for the two asymptotic limits, $\hat{s}^{2}\gg\lambda$
and $\hat{s}^{2}\ll\lambda,$ is

\begin{equation}
\lambda J_{2}(\hat{s}^{2},\lambda)\simeq\frac{\frac{1}{2\sqrt{2}}\left\{ 1-\frac{5}{4}\frac{\hat{s}^{2}}{\lambda}\left[1-\log\left(1+\sqrt{2}\right)\left(\frac{\hat{s}^{2}}{\lambda}\right)^{1/3}\right]\right\} }{1+\frac{5}{8\sqrt{2}}\left(\frac{\hat{s}^{2}}{\lambda}\right)^{7/3}},\label{eq:j2pade}
\end{equation}
see Fig. \eqref{fig:J2numvsan}. 

\begin{figure}
\includegraphics[scale=1.1]{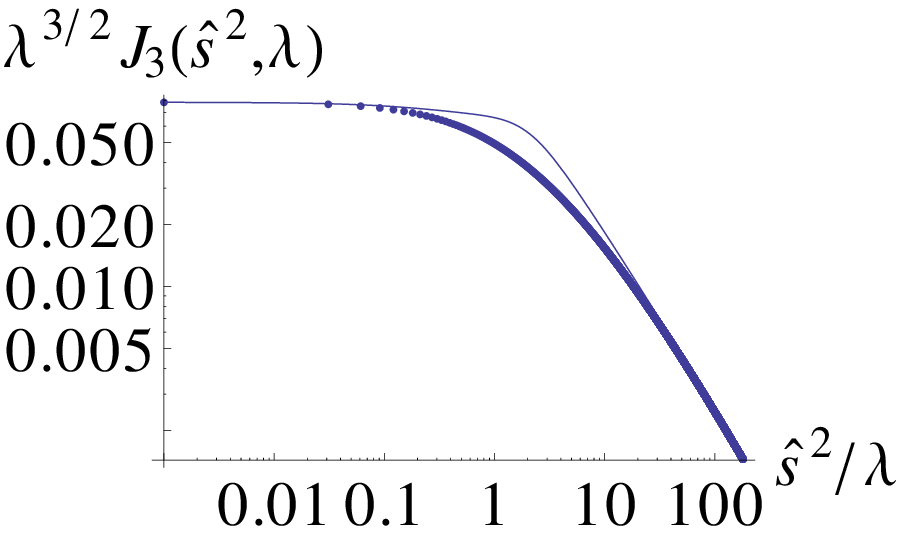}\caption{Comparison of the numerical solution (points) and the analytical solution
(line) of $J_{3}$ as defined in Eq. \eqref{eq:j3pade}.}

\label{fig:J3numvsan} 
\end{figure}

For the integral $J_{3},$ we have

\begin{equation}
\lambda^{3/2}J_{3}(\hat{s}^{2},\lambda)\simeq\sqrt{\frac{\pi}{2}}\frac{\frac{1}{8}-\frac{3}{32}\frac{\hat{s}^{2}}{\lambda}\left[1-\left(\frac{\hat{s}^{2}}{\lambda}\right)^{1/3}\right]}{1+\frac{3}{16}\left(\frac{\hat{s}^{2}}{\lambda}\right)^{7/3}},\label{eq:j3pade}
\end{equation}
see Fig. \eqref{fig:J3numvsan}. 

\begin{figure}
\includegraphics{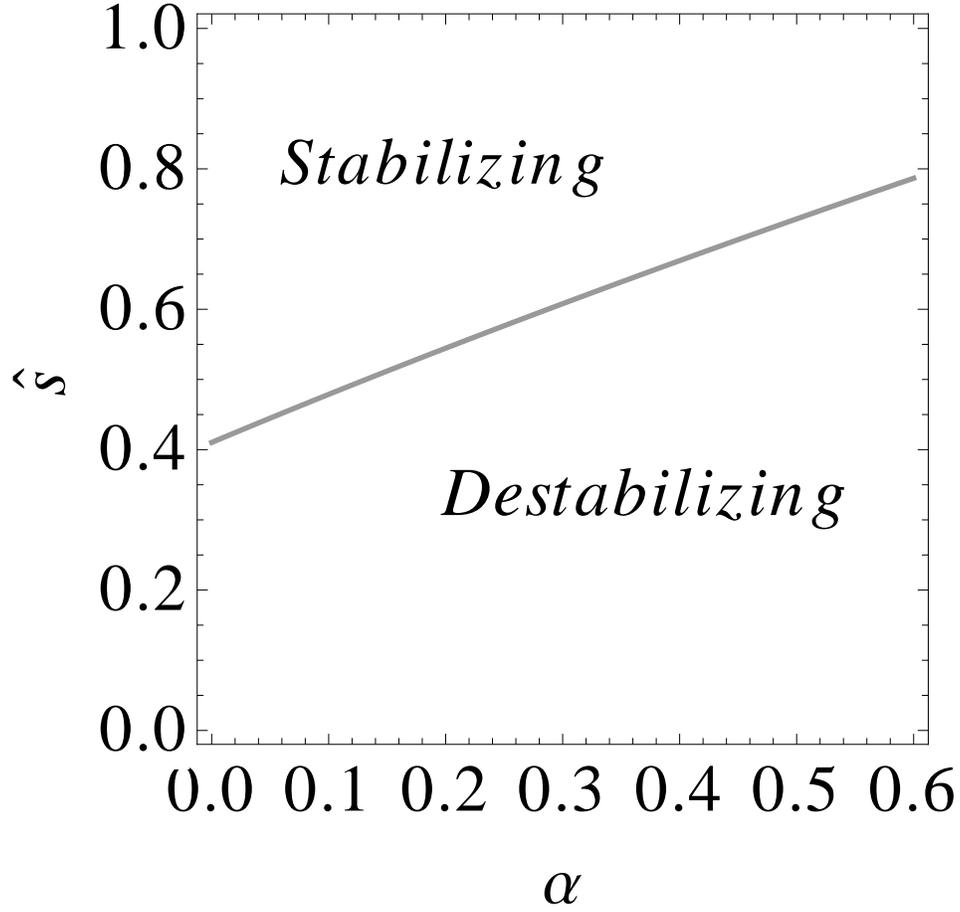}\caption{Contour plot of $\Im[\delta\omega]=0$ for an $\hat{s}-\alpha$ equilibrium
($a=0.5-\hat{s}+\alpha$). Here $\omega_{\kappa}/\omega_{*i}=0.04,$
$v_{thi}/(\sqrt{2}l_{\parallel}\omega_{*i})=10^{-2},$ $b_{0}=0.1,$
$\eta_{i}=10,$ $\tau=1,$ $\beta_{i}=10^{-4}.$ Toroidal electrostatic
drive: $\Im[\omega_{0}/\omega_{*i}]\gg\Re[\omega_{0}/\omega_{*i}],$
and $\Re[\lambda]>0.$}

\label{fig:EM_s_alpha_tor} 
\end{figure}

\begin{figure}
\includegraphics[scale=1.1]{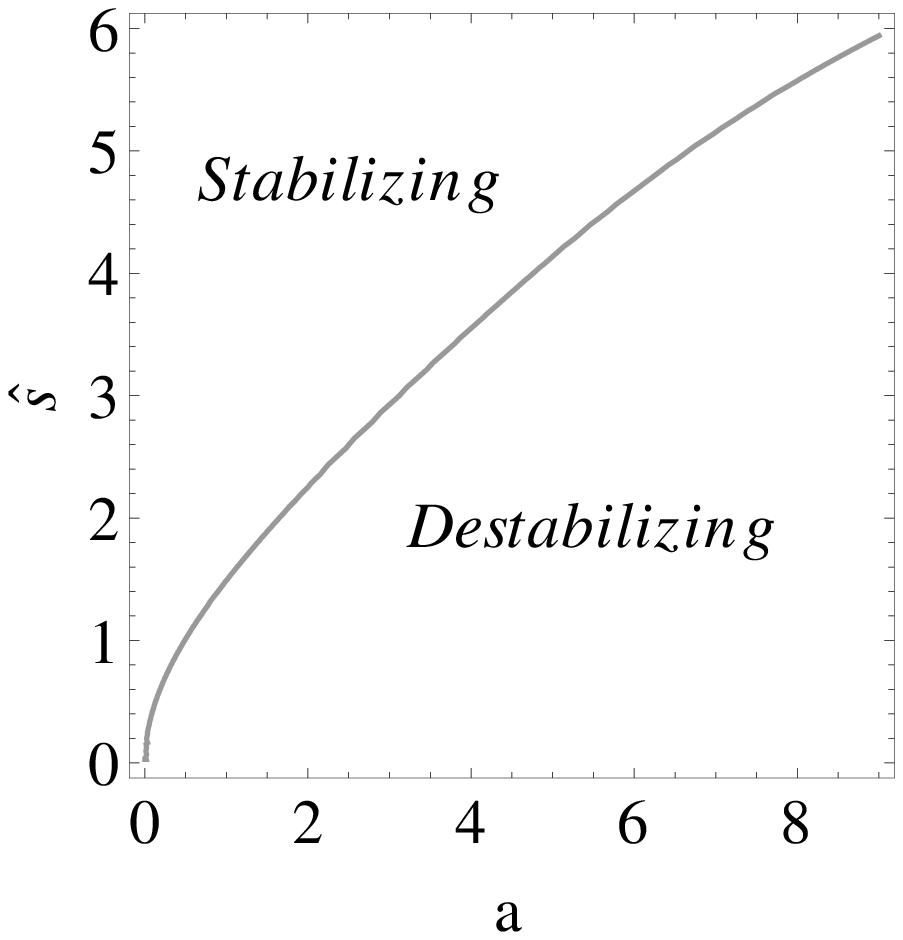}\caption{Contour plot of $\Im[\delta\omega]=0$ for a general equilibrium with
$a=0.5-\hat{s}+\alpha,$ and $\alpha=-R_{0}q^{2}\beta^{\prime}$ normalised
pressure grandient parameter of ideal MHD. Other parameters are as
in Fig. \eqref{fig:EM_s_alpha_tor}. }

\label{fig:EM_tor} 
\end{figure}

In Figs. \eqref{fig:EM_s_alpha_tor}-\eqref{fig:EM_tor}, we plot
the contour of $\Im[\delta\omega]=0$ for a Tokamak $\hat{s}-\alpha$
equilibrium \cite{CHT}, and for a general equilibrium with arbitrary
$a$ \cite{1.4868412}. We have introduced the familiar normalised
pressure grandient parameter of ideal MHD theory, $\alpha=-R_{0}q^{2}\beta^{\prime}.$
By expanding for small argument the poloidal dependence of the magnetic
drift frequency in an axisymmetric field, $\omega_{d}^{symm}\propto\cos\theta+\left[\hat{s}\theta-\alpha\sin\theta\right]\sin\theta,$
we find $a=0.5-\hat{s}+\alpha.$ We then use $\omega_{\kappa}/\omega_{*i}=0.04,$
$v_{thi}/(\sqrt{2}l_{\parallel}\omega_{*i})=10^{-2},$ $b_{0}=0.1,$
$\eta_{i}=10,$ $\tau=1,$ $\beta_{i}=10^{-4},$ and Eqs. (\ref{eq:jipade}-\ref{eq:j3pade}).
We checked a posteriori that for these parameters $\Re[\lambda]>0.$
In Fig. \eqref{fig:EM_s_alpha_tor} we see that, for the $\hat{s}-\alpha$
equilibium, the magnetic shear is stabilizing whereas, as expected,
$\alpha$ has a destabilizing effect. The critical $\alpha$ for destabilization
is a growing function of the local shear. For a generic equilibrium
{[}Fig. \eqref{fig:EM_tor}{]}, we find a critical length (for stabilization)
of the extent of the unfavorable curvature along the field, $\delta_{D}\sim l_{\parallel}a^{-1/2}.$
This quantity is a decreasing function of the magnetic shear. As in
the $\hat{s}\equiv0$ case of Sec. \eqref{sec:shearless}, an electron
contribution is the main cause of stabilization. However, now the
electron component of the parallel magnetic compressibility is subdominant,
since it does not depend on the local shear, $\hat{s,}$ while $J_{1}\sim\hat{s}^{-1}\gg1,$
for $\hat{s}\ll1.$ The dominant term $J_{1}$ is generated by the
first order correction to the parallel component of the magnetic potential,
$\psi^{(1)},$ calculated in Eq. \eqref{eq:psi1}, which is responsible
for electron parallel streaming, as evident from the electron solution
Eq. \eqref{eq:elsol}.

\section{\textup{Conclusions. }}

In the present work, we have revisited the problem of how curvature-driven
ITG instabilities are affected by finite plasma pressure. As is well
known, the latter affects both the equilibrium and the perturbed magnetic
drifts of the ions, and these effects partly cancel each other. If
the magnetic-field curvature is held constant while the electron +
ion pressure is increased, the equilibrium $\nabla B$-drift is reduced
in bad-curvature regions, see Eq. (6), which is stabilizing. On the
other hand, the finite {\em ion} pressure gradient also introduces
a new ${\bf B}\times\nabla\delta B_{\|}$ ion drift, which is destabilizing
by a mechanism identified in Fig. 1 and tends to cancel the stabilizing
effect of the ion pressure gradient (if the curvature $\boldsymbol{\kappa}$
is held constant). There remains, however, the stabilizing action
of the equilibrium {\em electron} pressure gradient, which stabilizes
the curvature-driven ITG mode at an electron beta of order $\beta_{e}\sim L_{T_{e}}/R$.
This scaling, heuristically derived in Sec. II, is confirmed quantitatively
by the solution of Eq. \eqref{eq:genshearless} and shown in Figs.
\eqref{Flo:6} and \eqref{Flo:7}. The general dispersion relation
in Eq. \eqref{eq:genshearless}, however, also captures the ion $\beta_{i}$
for the destabilization of ideal MHD modes, $\beta_{{\scriptscriptstyle MHD}}\sim L_{T_{i}}/(2q^{2}R)$.
The toroidal branch of the ITG can be completely stabilized for $\beta_{i}\gtrsim\beta_{{\scriptscriptstyle MHD}}.$
The solution of Eq. \eqref{eq:genshearless}, plotted in Fig \eqref{Flo:2-1},
shows such stabilization. Figures \eqref{Flo:6-1} and \eqref{Flo:7-1}
confirm the scaling of the critical beta for stabilization $\beta_{i}^{crit}\sim L_{T_{i}}/(2q^{2}R).$
The comparison of $\beta_{e}\sim L_{T_{e}}/R$ and $\beta_{{\scriptscriptstyle MHD}}\sim L_{T_{i}}/(2q^{2}R)$
determines which effect is more important in the electromagnetic stabilization
of the ITG mode. In a gyrokinetic code, this phenomenology is fully
accounted for only if the magnetic-field perturbation $\delta B_{\|}$
is included. In particular, in its absence, the destabilizing action
of the ${\bf B}\times\nabla\delta B_{\|}$ ion drift will be missed
and the code will tend to underestimate curvature-driven ITG instability. 

A third critical $\beta_{f}^{crit}$ for stabilization might be caused
by the presence of a fast particle species. We argue that the scaling
for $\beta_{f}^{crit}$ should be in qualitative agreement with \textbf{$\beta_{e}^{crit}\sim L_{T_{e}}/R,$}
due to some similarities in the response of a fast population and
electrons. Also in this case, a key role is played by the stabilizing
action of the equilibrium {\em fast particle} pressure gradient.

The results obtained from the local dispersion relation Eq. \eqref{eq:genshearless}
are valid when the magnetic shear and the finite extent (along the
field) of the bad-curvature region are negligible, unlike in a toroidal
device. When these are retained, we have shown that the effect of
a small plasma pressure gradient can be determined by perturbation
theory. Since the unperturbed (zero-$\beta$) operator is self-adjoint,
the amount of stabilization or destablization can be determined without
calculating the perturbed eigenfunctions. The resulting expression
\eqref{eq:deomegafin} is nevertheless complicated but predicts that
the extent of the unfavourable curvature along the magnetic field
needed for electromagnetic stabilization is a decreasing function
of the magnetic shear.

\bibliographystyle{iopart-num}  \bibliographystyle{iopart-num}
\bibliography{ZHCextended}

\section{Appendix}

As mentioned in the Introduction and at several places in the literature
\cite{Berk,0029-5515-20-11-011,1.3432117,1.3495976}, the destabilizing
effect of the ${\bf B}\times\nabla\delta B_{\|}$ drift is approximately
cancelled by the stabilizing influence of the finite-$\beta$ modification
of the equilibrium drift velocity. Mathematically, this cancellation
can be seen directly from the kinetic equation for the distribution
function $f=f_{0}+\delta f$, where the following combination of terms
appear in first order, 
\[
{\bf v}_{d}\cdot\nabla\delta f+\delta{\bf v}_{d}\cdot\nabla f_{0}.
\]
Substituting the expressions (\ref{eq:vdriftpert},\ref{eq:vdrift})
for ${\bf v}_{d}$ and $\delta{\bf v}_{d}$ from the Introduction
gives 
\[
{\bf v}_{d}\cdot\nabla\delta f+\delta{\bf v}_{d}\cdot\nabla f_{0}={\bf v}_{\kappa}\cdot\nabla\delta f-\frac{\mu_{0}v_{\perp}^{2}}{2\Omega B}{\bf b}\cdot\left(\nabla p\times\nabla\delta f-\nabla f_{0}\times\nabla\delta p_{\perp}\right).
\]
The terms within the brackets obviously have the tendency to cancel,
and indeed do so exactly when the divergence of the current is calculated,
which is effectively what is done in deriving Eq.~(\ref{eq:currdivbef-1}).
If we multiply by the charge, integrate over velocity space and sum
over all species $s$, these terms disappear: 
\[
\sum_{s}e_{s}\int\left({\bf v}_{ds}\cdot\nabla\delta f_{s}+\delta{\bf v}_{ds}\cdot\nabla f_{s0}\right)d^{3}v=\sum_{s}e_{s}\int{\bf v}_{\kappa s}\cdot\nabla\delta f_{s}\; d^{3}v.
\]

\end{document}